# Deep Learning for Automated Screening of Tuberculosis from Indian Chest X-rays: Analysis and Update

Anushikha Singh[1], Brejesh Lall[2], B.K. Panigrahi[2], Anjali Agrawal[3], Anurag Agrawal[4], Balamugesh Thangakunam[5], DJ Christopher[5]

[1]Bharti School of Telecommunications Technology & Management, Indian Institute of Technology Delhi, New Delhi 110016, India
[2]Department of Electrical Engineering, Indian Institute of Technology Delhi, New Delhi 110016, India
[3]Teleradiology Solutions, Civil Lines, Delhi 110054, India
[4]CSIR-Institute of Genomics and Integrative Biology, New Delhi 110025, India
[5]Department of Pulmonary medicine, Christian Medical College, Vellore - 632004, India
Anushikha.Singh@dbst.iitd.ac.in, brejesh@ee.iitd.ac.in, bijayaketan.panigrahi@gmail.com, anjali.agrawal@telradsol.com, a.agrawal@igib.in, drbalamugesh@yahoo.com, djchris@cmcvellore.ac.in

*Abstract*—**Background and Objective:** Tuberculosis (TB) is a significant public health issue and a leading cause of death worldwide. Millions of deaths can be averted by early diagnosis and successful treatment of TB patients. Automated diagnosis of TB holds vast potential to assist medical experts in expediting and improving its diagnosis, especially in developing countries like India, where there is a shortage of trained medical experts and radiologists. To date, several deep learning based methods for automated detection of TB from chest radiographs have been proposed. However, the performance of a few of these methods on the Indian chest radiograph data set has been suboptimal, possibly due to different texture of the lungs on chest radiographs of Indian subjects compared to other countries. Thus deep learning for accurate and automated diagnosis of TB on Indian datasets remains an important subject of research. **Methods:** The proposed work explores the performance of convolutional neural networks (CNNs) for the diagnosis of TB in Indian chest x-ray images. Three different pre-trained neural network models, AlexNet, GoogLenet, and ResNet are used to classify chest x-ray images into healthy or TB infected. The proposed approach does not require any pre-processing technique. Also, other works use pre-trained NNs as a tool for crafting features and then apply standard classification techniques. However, we attempt an end to end NN model based diagnosis of TB from chest x-rays. The proposed visualization tool can also be used by radiologists in the screening of large datasets. **Results:** The proposed method achieved 93.40% accuracy with 98.60% sensitivity to diagnose TB for the Indian population. **Conclusions:** The performance of the proposed method is also tested against techniques described in the literature. The proposed method outperforms the state of art on Indian and Shenzhen datasets.

*Index Terms*—Chest x-ray (CXR) image, Deep Learning, Tuberculosis, Artificial Intelligence.

## I. Introduction

Tuberculosis (TB) is an infectious disease and a leading cause of death worldwide. As per the World Health Organization (WHO) report 2019, approximately 10 million people were affected by TB across all countries and age groups [1]. It is caused by Mycobacterium tuberculosis which can affect the chest and other body sites [2]. As per WHO, it is estimated that the risk of developing TB is much higher among people affected by the human immunodeficiency virus - acquired immune deficiency syndrome (HIV-AIDS) [1-2]. People suffering from diabetes, under-nutrition, alcohol abuse and smoking are also at a higher risk of developing tuberculosis.

Chest x-ray (CXR) is a widespread and low-cost diagnostic tool used by medical experts for the screening of TB [3]. CXR contains a large amount of information which helps in the diagnosis of various lung diseases like TB, pneumonia, emphysema, and lung cancer [4]. Chest radiograph interpretation requires expertise and has a high inter- and intra-observer variability. Early diagnosis and treatment of tuberculosis are important to decrease the mortality rate and curtail spread of infection. Automated detection of TB from chest x-ray images is a very promising way to assist overburdened medical experts in screening a large population [5-6]. In literature, several works have been reported for automated diagnosis of TB from CXR broadly using three types of techniques, 1. Purely handcrafted, 2. Neural network (NN) based feature extraction followed by traditional classifiers, and 3. End to end NN based. We provide a brief overview explaining these methodologies.

### A. Handcrafted techniques for diagnosis of TB

Handcrafted techniques rely on texture to construct the features indicating the presence of TB. These techniques extract first order statistical and texture features [9, 17], shape abnormalities [12], wavelet based texture measure [8], features from Principal Component Analysis [10], fusion of different features [13-14] for classification of CXR into healthy or unhealthy/TB categories. For instance, Utkarsh et al. [15] considered novel features like height difference of lungs, lung bottom curvature and lower lung intensity variation for detection of pleural effusion.

All these handcrafted methods for detection of TB from chest x-rays provide insights, but are unable to compete with NN with regards to overall performance. In the current scenario, deep learning has wide applicability in the domain of medical image analysis. In many cases, it was observed that deep learning successfully replaces the process of handcrafted methods: feature extraction and disease classification [18]. CXR is a complex image and constructing a good set of features for the classification of normal and abnormal classes is challenging. Hence, deep learning is helpful and different deep learning based TB detection methods have been reported.

*B. NN based features extraction followed by traditional classifiers for diagnosis of TB*

This type of technique employs pre-trained convolutional neural networks as feature extractors which are further classified by a traditional classifier. Islam et al. [19] extracted features from different layers of pre-trained AlexNet [20], VggNet [21] and ResNet [22] architectures. Based on these features, traditional classifiers detect and localize abnormalities on chest x-rays. Lopes et al. [23] also used pre-trained GoogLenet, VggNet and ResNet networks for feature extraction from chest x-rays. These features were further fed to traditional classifiers for the detection of TB.

*C. End to end NN based techniques for diagnosis of TB:*

These techniques use both untrained and pre-trained CNN to classify the image into healthy or unhealthy classes. Hwang et al. [24] proposed the first method for deep learning based TB detection. They used AlexNet [20] based 9-layer architecture with a transfer learning strategy for classification of chest x-ray into normal and TB categories. Rahib et al. [25] detected multiple chest diseases from chest x-ray using Back propagation NNs and Competitive NNs with supervised and unsupervised learning, respectively. Lakhani et al. [26] used untrained and pre-trained networks, AlexNet and GoogLenet [27] for the detection of TB from chest x-rays. Rajpurkar et al. [28] presented a dense convolutional network [29] for the detection and localization of 14 disease labels from chest x-rays. Recently Pasa et al. [30] designed a simple 5 blocks NN architecture for TB detection and achieved comparable performance with the advantage of fewer parameters.

Various authors reported that the automated detection of TB from chest x-rays is challenging. The main reason for this is the spread of TB findings over the entire region of the lung. Further, these findings may overlap with those for other pulmonary diseases. All reported works on NN models show good performance but highlight two important tenets for their success: Firstly, NN learning schemes are highly dependent on datasets and hence high quality data is critical for a successful model, and secondly, a need for higher accuracy in the case of medical diagnosis. The published models are tested on standard datasets and that does not guarantee their performance on the Indian dataset. Hence, there is a need to evaluate the performance of NNs on Indian chest x-rays.

This work presents the efficient use of CNNs for the detection of TB from chest x-ray images.
The salient features of our work are:
- We attempt end to end deep convolutional neural networks to diagnose TB for the Indian population as the texture of Indian lungs is different from those of subjects from other countries, like the USA, Africa, and China, on whom the existing models have been built [34].
- The TB detection methods used in the literature apply NN as a tool for feature extraction which is further classified by a traditional classifier. However, the performance of a few of these methods on the Indian chest radiograph data set has been suboptimal in comparison to standard datasets. Our work achieves high accuracy and sensitivity for both in-house Indian and public Shenzhen dataset.
- Unlike the deep learning-based techniques reported in the literature, the proposed work does not involve any pre-processing methods like lung segmentation or rib suppression.
- We produce heat maps for visualization of network prediction and to identify the most considerable area by the NN model for the detection of TB. We test the visualization capabilities of the NN model and discuss them from a radiologist perspective.
- The proposed work is evaluated on Indian chest x-rays, including mild, moderate, and severe affected lung fields due to TB which makes it applicable for TB diagnosis in real-world scenarios.

The manuscript is structured as follows: Section 2 presents a detailed methodology that includes NN architectures with their training, data augmentation and tuning of hyper-parameters involved. Section 3 describes the datasets used for training and testing the proposed method. Section 4 presents experimental results for the detection of TB and visualization of TB findings on the chest x-ray images. It also includes a comparative study of the proposed method with existent literature and discussion of the results. Finally, section 5 includes the conclusions.

II. METHODOLOGY

The objective of this work is to explore the use of end to end NNs for detection of TB from chest x-ray images. Pre-trained CNNs AlexNet, GoogLenet and ResNet were used for classification of chest x-ray images into two classes: normal or TB. We use pre-trained networks trained on the ImageNet database that contains 1.2 million color images and 1000 classes. Each architecture has some unique characteristics which are discussed below:

*A. NN architectures used*

*AlexNet:* Alex Krizhevsky et al. [20] proposed AlexNet (winner of 2012 ImageNet LSVRC-2012 competition) network which involves 60 million parameters and eight layers including five convolutional and three fully-connected layers. Fig. 1 shows the architecture of AlexNet model. The network works with Rectified linear units (ReLUs) non-linearity for faster training. If input is $x$ and the corresponding neuron's output is $f(x)$ then ReLUs non-linearity is defined as:

$$f(x) = \text{maximum}(0, x) \quad \text{(i)}$$

In this model, dropout regularization is used to reduce overfitting in the fully connected layers and stochastic gradient descent as a momentum optimizer.
The weight ($w$) update rule is:

$$v_{i+1} = 0.9 \cdot v_i - 0.0005 \cdot \varepsilon \cdot w_i - \varepsilon \cdot \left(\frac{\partial L}{\partial w}\bigg|_{w_i}\right)_{D_i} \quad \text{(ii)}$$

Where *i* is the iteration index, *v* is the momentum variable and Ɛ represents learning rate. $\left\langle \frac{\partial L}{\partial w}|_{w_i}\right\rangle_{D_i}$ is the average, over $i^{th}$ batch $D_i$ of the derivative of the objective function with respect to *w*, evaluated at $w_i$. The architecture draws initial weights from a zero-mean Gaussian distribution with std. deviation of 0.01. The loss function for this network is cross entropy which needs to be minimized. For the binary classification, the cross entropy loss function (*L*) is:

$$L= -(y \log (p) + (1-y) \log (1-p)) \quad \text{(iii)}$$

Where *y* is true class label and *p* is predicted probability.

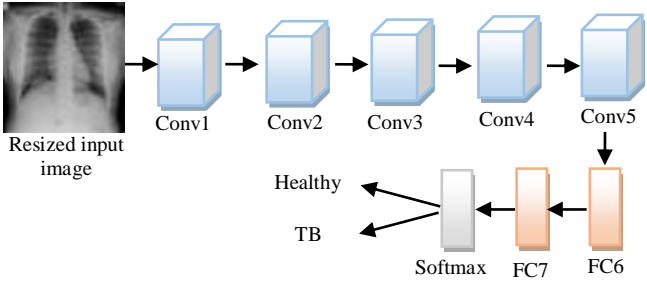

Fig. 1.The AlexNet architecture [20].

*GoogLenet:* Szegedy et al. [27] developed a 22-layer deep network GoogLenet which is the winner of ILSVRC – 2014 competition. This network involves 22 layers and 7 million parameters which are very less in number as compared to those in AlexNet. The use of inception module with dimension reduction makes this network computationally efficient. All the convolutional layers including those inside the inception module use rectified linear activation to train the network faster. Similar to the AlexNet model, this network also uses dropout regularization to reduce overfitting and stochastic gradient descent as a momentum optimizer. The architecture of GoogLenet network and inception module is shown in Fig. 2 and Fig. 2a respectively.

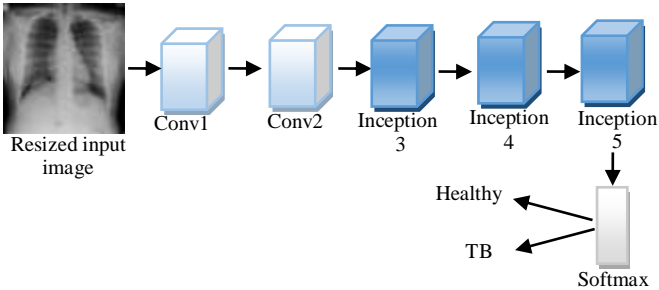

Fig. 2.The GoogLenet architecture [23].

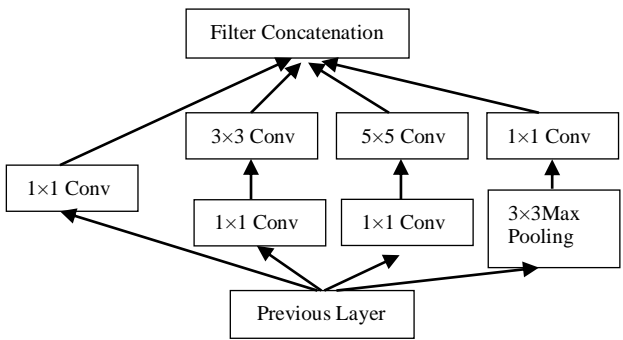

Fig. 2 (a).The Inception module [23].

*ResNet:* The third neural network used in this work is ResNet [22] (winner of ILSVRC & COCO 2015 competitions) and its different versions which are deeper than AlexNet and GoogLenet. The proposed work uses three different versions of Resnet: Resnet 18, ResNet 50, ResNet 101 which involves 18, 50, 101 layers and 11.7, 25.6, 44.6 million parameters respectively. In traditional NN architectures, the performance of neural network diminishes as we make the network deeper. The reason is the vanishing gradient issue which means as the number of layer increase, the gradients in layers close to input become unusually small. ResNet introduces a deep residual learning framework to address the degradation issue. The residual connection exploits the use of skip layer as shown in Fig. 3.

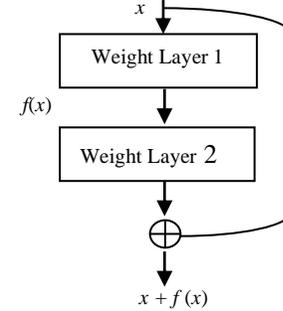

Fig. 3.The deep residual learning framework [26].

If *x* and *f(x)* is the input and output of each layer, then
In the case of standard network: *y = f(x)*     (iv)
However, for the residual network: *y = x + f(x)*     (v)
where y is the final output.

ResNet is constructed by stacking many such residual modules together. The ResNet network uses batch normalization layer between convolutional and ReLUs layers for faster training and to reduce the sensitivity to network initialization.

*B. Incremental training:*

The pre-trained NNs used in the proposed work were trained on the ImageNet database. The NNs were not trained on chest x-rays so incremental training on our specific dataset is required. The dataset used for incremental training contains a total of 550 images in which 216 images belong to healthy subjects and the remaining 334 images belong to TB patients. This dataset is subdivided into 3 sets: training, validation, and testing. 50% of total images were used for incremental training of NNs and remaining 50% (25%: validation, 25%: testing) image were used for validation and testing of NNs.

*C. Data Augmentation:*

The pre-trained networks require input images of a particular size so all training images were resized. Additional augmentation: randomly flipping along the vertical axis and randomly translation up to 30 pixels horizontally and vertically was done to increase the amount of training data. Data augmentation is also required to prevent the NNs from overfitting and learning the precise details of training images.

*D. Tuning of Hyper-parameters:*

The stochastic gradient descent is used for training with a momentum optimizer. The other hyper-parameters are mini-batch size which is considered as 10, the number of epochs is 20, and the initial learning rate is selected as 0.0001.

## III. DATASET

The proposed work is to explore the use of CNNs for the detection of TB from Indian chest x-rays. To compare the performance of the proposed work with the existing literature, the proposed work is also tested on a publicly available Shenzhen dataset. The details of both Indian and Shenzhen datasets are given below. The primary objective of this work is to propose a fast and accurate automated TB detection method for the Indian dataset. The use of a dataset representing a different demography has been done to provide a comparison of performance.

### A. Indian Dataset:

The postero-anterior CXRs were collected from Christian Medical College, Vellore, India. The dataset contains a total of 550 CXRs in which 216 CXRs belong to healthy subjects and the remaining 334 CXRs were from confirmed TB patients, between 2014-2017. TB was diagnosed in patients with clinical symptoms suggestive of TB, confirmed by Xpert MRB/RIF and/or mycobacterial culture. The age of the healthy subjects was 46.4±15.2 (mean±std. deviation) and 55.6% were males. The age of the tuberculosis patients was 46.6±16.6 (mean±std. deviation) and 69.8% were males. The TB CXRs were performed at the time of diagnosis and before or within 1 week of the initiation of anti-TB treatment. The TB CXRs belongs to mild, moderate and severe patients of active pulmonary tuberculosis. The CXRs of patients with other pulmonary diseases but not TB was excluded from the final database used in the proposed work. The digital X-rays were stored in hospital PACS system. All the CXRs were annotated and interpreted by an experienced pulmonologist and radiologist. The format of all CXRs is JPEG and the images are of varying resolution hovering in the range of 2000 × 2000. Table 1 presents demographic characteristics of the subjects. Fig. 4 shows four sample images of chest x-ray each of healthy and TB affected from the Indian dataset.

TABLE I
DEMOGRAPHIC CHARACTERISTICS OF INDIAN DATASET

| Demographic Parameters | Healthy cases | Tuberculosis cases |
|---|---|---|
| Number of patients | 216 | 334 |
| Age in years (mean ± std. deviation) | 46.4±15.2 | 46.6±16.6 |
| Gender ratio (male:female) | 55.6 : 44.4 | 69.8 : 30.2 |

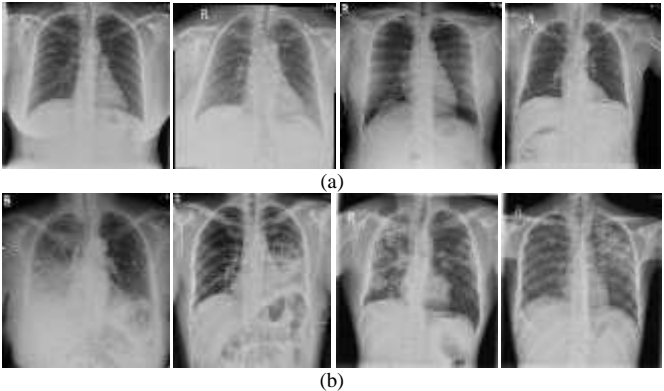

Fig. 4. Four Sample Images of Indian dataset (a) Healthy (b) TB affected.

### B. Shenzhen Dataset:

The images of Shenzhen dataset [31] were collected by the National Library of Medicine, Maryland, USA with the collaboration of Shenzhen No.3 People's Hospital, and Guangdong Medical College, Shenzhen, China. This dataset contains 662 chest X-ray images, including 326 belonging to healthy subjects and the remaining 336 belonging to TB patients. All these images are of different resolutions approximately 3000×3000 in terms of pixels and in JPEG format. Fig. 5 shows four samples of chest x-ray images each of healthy and TB affected from the Shenzhen dataset.

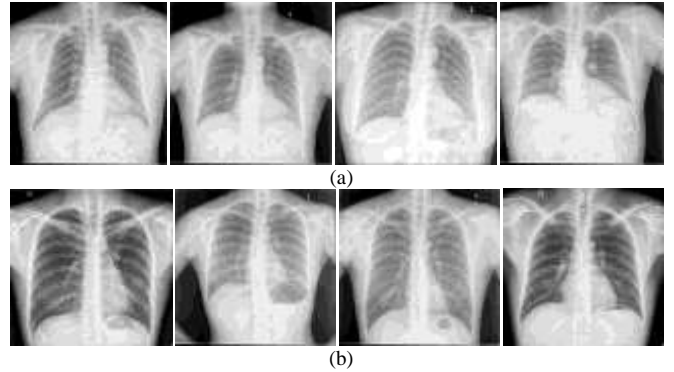

Fig. 5. Four Sample Images of Shenzhen dataset (a) Healthy (b) TB affected.

## IV. RESULTS AND DISCUSSION

Experiments were carried out to evaluate and validate the performance of the proposed method. Results were obtained by tuning the hyper-parameters to increase the performance and this section reports only the most accurate results. The proposed method is tested on 136 images (25% of total available images) of the Indian dataset in which 66 images are of healthy individuals and the remaining 70 images are those of TB patients.

### A. Performance Evaluation Metrics:

The proposed method is quantitatively evaluated by using the three well-known metrics:

$$\text{Sensitivity} = \frac{TP}{TP + FN} \quad \text{(vi)}$$

$$\text{Specificity} = \frac{TN}{TN + FP} \quad \text{(vii)}$$

$$\text{Accuracy} = \frac{TP + TN}{TP + TN + FP + FN} \quad \text{(viii)}$$

where true positive (TP) is the total number of TB images correctly identified as TB, true negative (TN) is the total number of healthy images correctly identified as healthy, false positive (FP) is the total of healthy images misclassified as TB and false negative (FN) presents the total number of TB images misclassified as Healthy.

Table 2 presents the performance of CNNs in terms of TP, TN, FP, and FN. TP and TN are the numbers of TB and healthy images classified correctly so these values should be closer to 100%. FP and FN are the measures for misclassification in terms of the number of images classified incorrectly. As per the results reported in Table 2, AlexNet and ResNet18 correctly classify all TB images as TB except one image. ResNet101 correctly classify 62 out of 66 healthy images as healthy.

TABLE II
CLASSIFICATION RESULTS: TP, TN, FP & FN

| Neural Network | TP | TN | FP | FN |
|---|---|---|---|---|
| | Correct Classification | | Misclassification | |
| AlexNet | 69/70 | 55/66 | 11/66 | 1/70 |
| GoogLenet | 65/70 | 56/66 | 10/66 | 5/70 |
| ResNet 18 | 69/70 | 58/66 | 8/66 | 1/70 |
| ResNet 50 | 63/70 | 61/66 | 5/66 | 7/70 |
| ResNet 101 | 65/70 | 62/66 | 4/66 | 5/70 |

Table 3 comprises results for all CNNs in terms of sensitivity, specificity & accuracy. The highest accuracy 93.40% for the detection of TB is provided by both AlexNet and ResNet model. In the case of disease diagnosis, sensitivity is a very important metric which should be high as a missed diagnosis of TB may have serious ramifications for both the individual and the community. In our case, AlexNet and ResNet18 model provides 98.60% sensitivity which is reasonably good. Only one image of a TB patient is misclassified as healthy and the rest are correctly classified. For specificity, the ResNet101 gives the best performance at 93.93%. Table 4 presents that the validation accuracy for different CNNs is more than 87% which means that we can expect our trained model to perform well on the new datasets.

TABLE III
CLASSIFICATION RESULTS: SENSITIVITY, SPECIFICITY, & ACCURACY (IN %).

| Neural Network | Sensitivity | Specificity | Accuracy |
|---|---|---|---|
| AlexNet | 98.60 | 83.30 | 91.18 |
| GoogLenet | 92.90 | 84.80 | 88.97 |
| ResNet 18 | 98.60 | 87.87 | 93.38 |
| ResNet 50 | 90.00 | 92.42 | 91.18 |
| ResNet 101 | 92.90 | 93.93 | 93.40 |

TABLE IV
THE VALIDATION ACCURACY (IN %)

| Neural Network | AlexNet | GoogLenet | ResNet18 | ResNet 50 | ResNet 101 |
|---|---|---|---|---|---|
| Validation Accuracy | 89.52 | 92.74 | 87.90 | 91.94 | 87.91 |

### B. Analysis of Receiver Operating Characteristics:

Receiver Operating Characteristics (ROC) curve [32] is the most popular performance measurement tool for classification problems. The ROC is a probability curve, plotted with sensitivity against 1-specificity at various threshold settings with sensitivity on the $y$-axis and specificity on the $x$-axis. The Area under curve (AUC) [33] is defined as an area under the ROC curve which represents the measure of separability between two classes. The highest value of AUC is 1 which means that the model has the highest possible level of capability to separate two classes. The worst value of AUC is 0 which means that the model is not able to separate those classes. We also performed the ROC-AUC analysis in our case and compared the different CNN architectures. Fig. 6 presents the ROC curves for AlexNet, GoogLenet, ResNet18, ResNet50, and ResNet101 networks. Fig. 6 shows that the performance of the CNNs is comparable for the classification of healthy and TB affected chest x-rays. Table 5 presents corresponding AUC values for testing and validation data separately which are greater than 0.95 for all the considered CNNs. The highest value of AUC is 0.98 which is obtained by both AlexNet and ResNet50 networks. The high value of AUC indicates that the proposed method works well for the diagnosis of TB from Indian chest x-rays.

TABLE V
THE AUC VALUES: VALIDATION AND TESTING

| Neural Network | AlexNet | GoogLenet | ResNet18 | ResNet 50 | ResNet 101 |
|---|---|---|---|---|---|
| Validation AUC | 0.9554 | 0.9789 | 0.9502 | 0.9842 | 0.9547 |
| Testing AUC | 0.9851 | 0.9721 | 0.9801 | 0.9848 | 0.9764 |

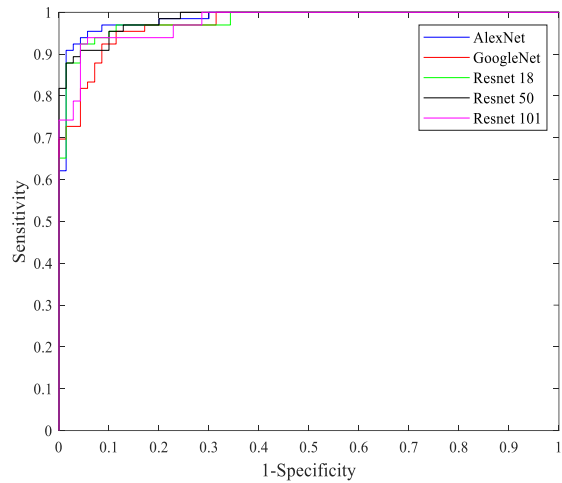

Fig. 6. Comparison of ROC curves for AlexNet, GoogLenet, ResNet18, ResNet50, and Resnet101 networks.

### C. Visualization of Detected Abnormalities in Chest X-ray:

The proposed work aims to develop a method that can assist the radiologist in the automated screening of TB from chest x-rays. From a radiologist perspective, Deep NNs are black boxes that do not provide any way to figure out which area of the input image to the model was accountable for the decision or what the model has learned from the input. There is no clue when this model fails and misclassifies the input image. Keeping this in mind, our work interprets the model prediction by generating heat maps to visualize the part of input chest x-ray most indicative of the TB manifestation. These visualization techniques help us understand the network and may also be useful as an approximate visual diagnosis for presentation to radiologists. We use class activation mapping (CAM) which would assist radiologists to check which part of the input image is responsible for correct classification or in case of an incorrect prediction [35]. To generate the heat map, the input image is fed to an end to end trained NNs. The feature map is extracted from the strongest activated channel of the final convolutional layer. The AlexNet model is preferred for visualization of abnormalities as it is a relatively shallow network as compared to the others. The strongest activation is selected from the last convolutional layer (5th convolution layer) and feature maps are extracted to generate the heat map. The selected feature map is overlaid on the input chest x-ray image to highlight the areas where symptoms of TB are present.

Fig. 7 shows four sample images of TB affected chest x-rays with their corresponding heat map generated using the method described above. The left image and right image of Fig. 7 presents the input image fed to the neural network and overlaid with corresponding heat map respectively. The generated heat map highlights the area of the image which is most likely to contain TB markers for making a classification decision. This visualization method helps us to understand the model and may also be useful as an approximate visual diagnosis for presentation to radiologists.

Based on results shown in Table 3, the AlexNet model provides 98.60% sensitivity which indicates that only one image of a TB patient is misclassified as healthy and the remaining TB images are correctly classified. Fig. 8 shows the visualization result for that one particular TB affected image which was misclassified as healthy. As per the radiologist, signs of TB were present in the upper and mid-zone of the left lung in that image, but the proposed method failed to detect it. The heat map highlights some areas in the background and some areas in the middle zone of the left lung which is not significant for consideration of TB. In the right image of figures 7 & 8, the class activation map shows which regions of the input image contribute the most to the predicted class TB. Red regions contribute the most. All these results were validated by a radiologist from Teleradiology Solutions, New Delhi, India.

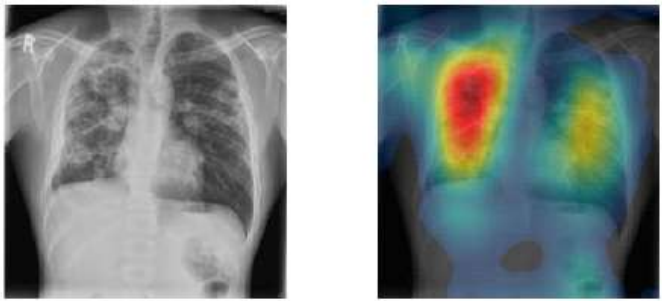

(a) Chest x-ray of a patient where symptoms of TB are present in both right and left lung area (left image). Our NN model correctly detects TB and also identifies area in the image most indicative of pathology (right image).

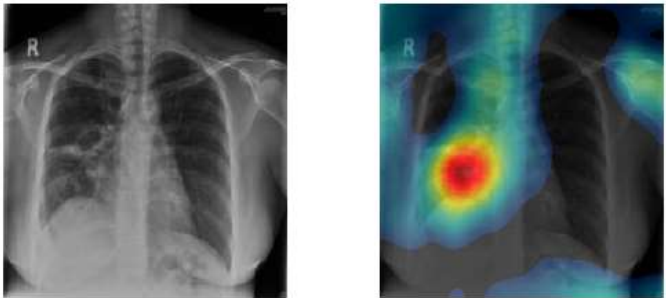

(b) Chest x-ray of a patient where symptoms of TB are present in both right lung area (left image). The proposed NN model correctly localizes the diseased area and classify image as TB (right image).

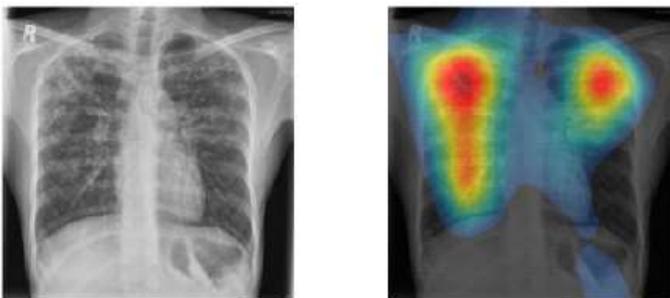

(c) Chest x-ray of a patient where symptoms of TB are present in the entire right lung along with upper and mid zone of the left lung (left image). Our model considers diseased area to make correct classification decision (right image).

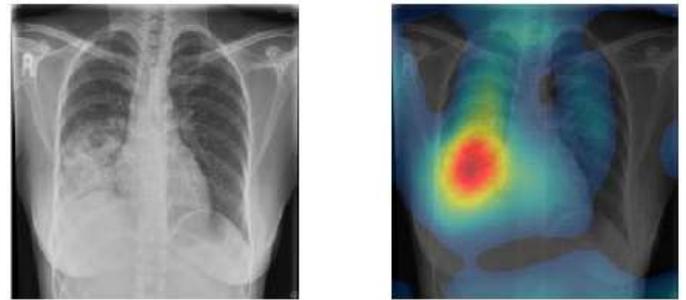

(d) Chest x-ray of a patient where symptoms of TB are present in right lung area (left image). Our NN model correctly localizes the diseased area and classify image as TB (right image).

Fig. 7. Visualization of abnormalities identified in TB affected chest x-rays.

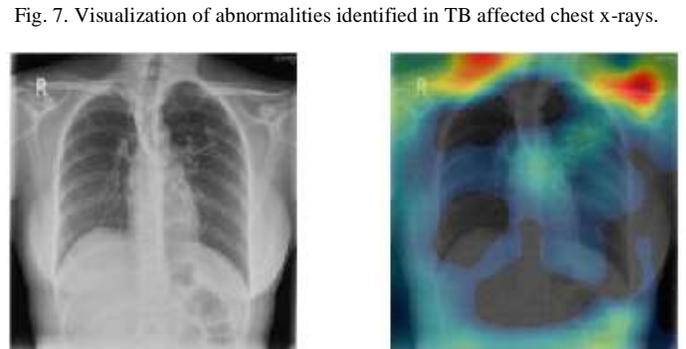

Fig. 8. The visualization results on one TB affected image which was misclassified as healthy by AlexNet model. The heat map indicates the background area most responsible for incorrect classification.

### D. Comparison with Current State of the Art: Indian Dataset

In the literature [19 & 23], different CNNs are employed as a feature extractor from chest x-ray images and these features are classified by traditional classifiers for detection of TB. We conducted similar experiments on the Indian dataset using a feature based approach [23] and compared our results against it. Table 6 and Table 7 present results reported using a feature based approach on the Indian dataset and those are compared with the results achieved by our proposed work. Table 6 reports accuracy and AUC values. The highest accuracy of 92.60% is achieved by the AlexNet model for the feature based approach. Our proposed method obtained 93.40% accuracy using the ResNet model which is better than a feature based approach. Similarly, in terms of AUC, our proposed method achieved superior value in comparison with feature based approach. Since in the case of medical image classification, sensitivity and specificity are very important aspects of performance measurement, Table 7 presents those results. The maximum sensitivity is 94.30% using a feature based approach and it increases to 98.60% using our proposed method. The maximum value of specificity is 93.90% using our proposed method while it drops down 90.90% using a feature based approach.

TABLE VI
COMPARISON OF OUR WORK WITH FEATURE BASED APPROACH IN TERMS OF ACCURACY AND AUC ON INDIAN DATASET

| Neural Network | Feature Based Approach [23] | | Our Work | |
|---|---|---|---|---|
| | Accuracy (in %) | AUC | Accuracy (in %) | AUC |
| AlexNet | 92.60 | 0.9794 | 91.18 | 0.9851 |
| GoogLenet | 88.20 | 0.9444 | 88.97 | 0.9721 |
| ResNet 18 | 89.00 | 0.9563 | 93.38 | 0.9801 |
| ResNet 50 | 89.67 | 0.9656 | 91.18 | 0.9848 |
| ResNet 101 | 89.70 | 0.9621 | 93.40 | 0.9764 |

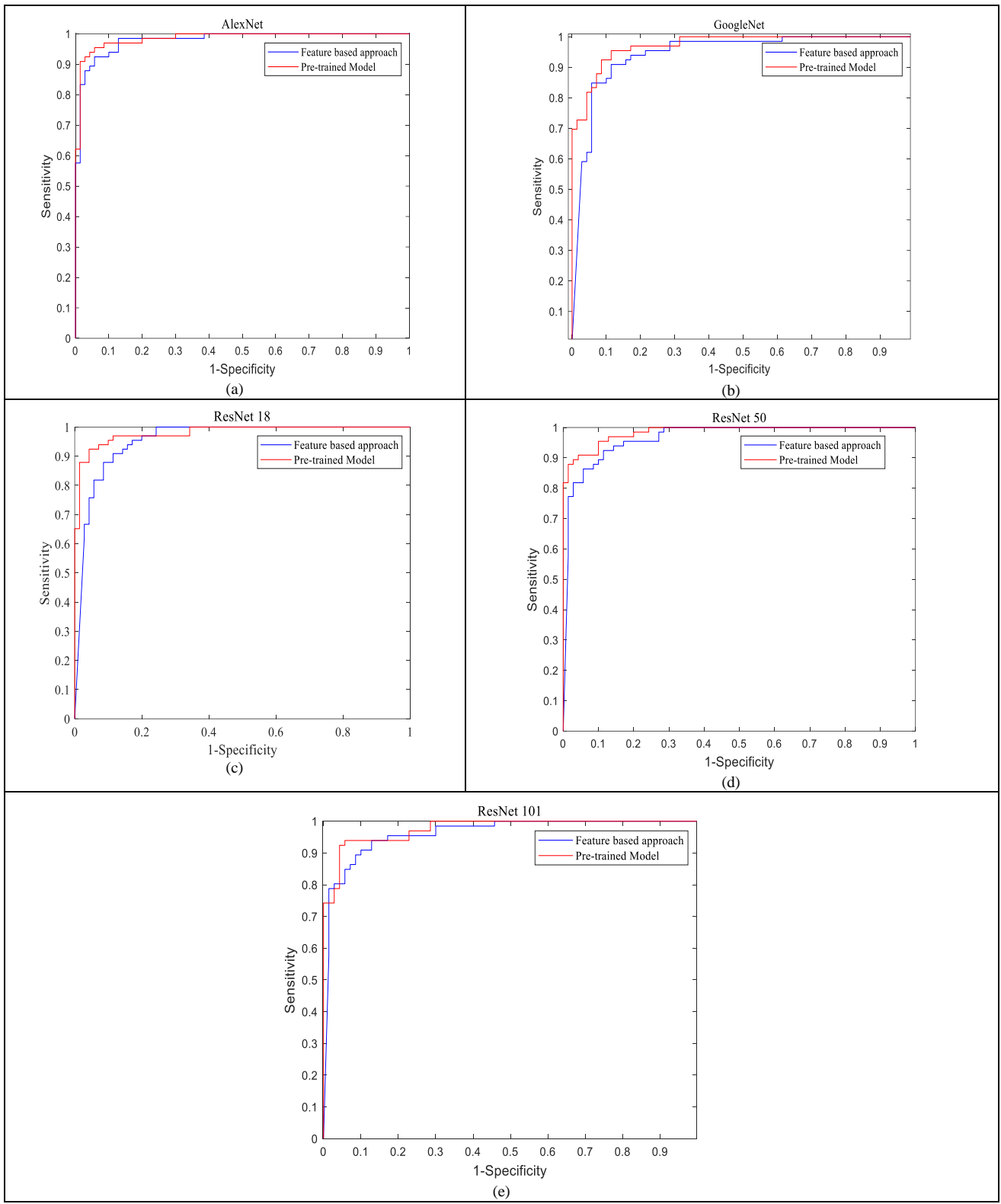

Fig. 9. Comparison of ROC curves for different CNNs. (a) AlexNet, (b) GoogLenet, (c) ResNet18, (d) ResNet50, and (e) Resnet101 models.

TABLE VII
COMPARISON OF OUR WORK WITH FEATURE BASED APPROACH IN TERMS OF
SENSITIVITY AND SPECIFICITY (IN %) ON INDIAN DATASET

| Neural Network | Feature Based Approach [23] | | Our Work | |
|---|---|---|---|---|
| | Sensitivity | Specificity | Sensitivity | Specificity |
| AlexNet | 94.30 | 90.90 | 98.60 | 83.30 |
| GoogLenet | 88.60 | 87.90 | 92.90 | 84.80 |
| ResNet 18 | 88.60 | 89.40 | 98.60 | 87.90 |
| ResNet 50 | 91.40 | 87.90 | 90.00 | 92.40 |
| ResNet101 | 94.30 | 84.80 | 92.90 | 93.90 |

Fig. 9 shows a comparison of ROC curves obtained using a feature based approach and the proposed work for different CNNs. It is clear from the five ROC curves that the proposed method performs better in comparison to a feature based approach. All these experiments were done on the same Indian dataset for a fair comparison and it was observed that the proposed work gives superior results for Indian images.

*E. Performance Evaluation on Non-Indian Dataset:*

To evaluate the generalization capability of the proposed technique, we applied it to a standard non-Indian dataset and compared the results with those obtained by other methods. It is interesting to note that the methods we compared against are tuned for the chosen dataset whereas the proposed method has not been similarly tuned. Despite this disadvantage, the proposed technique outperforms all other reported techniques except one which it approximately matches. Table 8 and Table 9 shows the testing and validation results achieved by the proposed method on the standard Shenzhen dataset [31]. The AlexNet model obtained the highest accuracy of 87% with an AUC value of 0.9283 which is better than the other models used in this work.

TABLE VIII
TESTING RESULTS ON SHENZHEN DATASET: ACCURACY & AUC.

| Neural Network | Accuracy (in %) | AUC |
|---|---|---|
| AlexNet | 87.00 | 0.9283 |
| GoogLenet | 79.50 | 0.8921 |
| ResNet 18 | 85.50 | 0.9103 |
| ResNet 50 | 78.00 | 0.9147 |
| ResNet 101 | 84.50 | 0.9129 |

TABLE IX
VALIDATION RESULTS ON SHENZHEN DATASET: ACCURACY & AUC.

| Neural Network | Accuracy (in %) | AUC |
|---|---|---|
| AlexNet | 88.49 | 0.9590 |
| GoogLenet | 81.29 | 0.9072 |
| ResNet 18 | 84.89 | 0.9188 |
| ResNet 50 | 85.61 | 0.9575 |
| ResNet 101 | 84.17 | 0.9085 |

Table 10 presents comparative results obtained by the proposed technique with similar work on TB detection reported in the literature. This comparison is done with deep learning based methods that report their results on the Shenzhen dataset. Table 8shows that the proposed work achieved 87% accuracy for Shenzhen dataset which is higher than the works reported in [19], [27], [30] and comparable with Islam et al.'s work [24]. The survey paper [18] reveals that deep learning based methods consistently outperform handcrafted methods of TB detection, hence, the proposed work is not compared against these methods.

Lakhani et al. [22] used NNs for detection of TB from chest x-rays but the work is not included in the comparison Table 8. This study is excluded because the dataset used in their work combines images of four different datasets including Shenzhen, Montgomery, images from Belarus Tuberculosis Portal and Thomas Jefferson University Hospital, Philadelphia. Since the results on Shenzhen dataset are not reported separately, a fair comparison is not possible. All other deep learning based published works are included in Table 8 for comparison. Table 8 shows that our proposed technique is superior or at par with most others. In the case of TB detection from Indian CxR images, the proposed work achieved the highest accuracy of 93.40%.

As suggested in literature and including our work that use deep learning for TB detection, the images are resized to meet the requirement of the respective models before analysis by the NNs. This would help meet several parameters and layers inherent to these networks. On the surface, it seems that high resolution images may improve the performance of the NNs, but this would also increase the computational complexity and require a more robust platform. Large computation time (approx. 20 minutes using Intel i7 CPU, 3.60 GHz) required for incremental training of neural networks is another drawback of our study. These challenges would need to be addressed in the future.

TABLE X
COMPARISON OF THE PROPOSED WORK WITH LITERATURE ON SHENZHEN DATASET.

| Literature | Method/Model | Performance | |
|---|---|---|---|
| | | Accuracy | AUC |
| Hwang et al. [24] | AlexNet based 9 layer architecture Transfer learning strategy | 83.70 % | 0.9260 |
| Islam et al. [19] | AlexNet, VggNet, ResNet as a feature extractor & traditional classifiers | 88.00 % | 0.9100 |
| Lopes et al. [23] | GoogLenet, VggNet, ResNet for feature extraction & SVM Classification | 84.70 % | 0.9040 |
| Pasa et al. [30] | 5 convolution blocks network | 84.40 % | 0.9000 |
| The proposed work | Pre-trained CNNs: AlexNet, GoogLenet, ResNet Transfer learning strategy | 87.00 % | 0.9283 |

V. CONCLUSION

This paper proposes an efficient use of CNNs for automated screening of TB from chest x-ray images. Three different end to end NN models AlexNet, GoogLenet, ResNet were considered for the classification of healthy and TB affected chest x-rays. The networks were pre-trained on a large ImageNet database prior to incremental training on chest x-rays. Further, heat maps were generated from feature maps associated with the strongest activated channel of the final convolutional layer as an aid for medical experts. Analysis revealed that the proposed method achieved high accuracy for the classification of healthy and TB affected chest x-rays. It had high generalizability demonstrated by the strong performance achieved for another standard publicly available dataset, namely, Shenzhen. Among the CNN models used in the proposed work, the best accuracy of 93.40% is achieved by ResNet architecture. In the case of the Shenzhen dataset, the AlexNet model was the most accurate at 87%. A possible explanation is that a deep architecture is slow to respond to new datasets. ResNet with its deep architecture is presumably not required for a database with less data variability. These results also indicate that the Indian dataset has a greater variability in comparison to the Shenzhen dataset. Based on these results, it is believed that the proposed work is valuable for automated screening of TB from CxR for Indian databases.

ACKNOWLEDGEMENT

The study was administered under CRDF Global grant (OISE-17-62923), under the aegis of the REPORT India consortium. The consortium is supported by the Department of Biotechnology, Ministry of Science and Technology and the National Institute for Allergy and Infectious Diseases and National Institutes of Health (USA). DJ Christopher, and Balamugesh Thangakunam collected the Indian patients' chest X-ray images, interpreted, correlated with the clinical diagnosis & classified them. Dr. Anurag Agrawal, CSIR-IGIB, India, for coordinating the multi-institutional team, and critical inputs, and Dr. Anjali Agrawal, Teleradiology Solutions collected and labeled the images.